# Invited Article: 4D Printing as a New Paradigm for Manufacturing with Minimum Energy Consumption

Farhang Momeni [1,*] and Jun Ni [1]

[1] Department of Mechanical Engineering, University of Michigan, Ann Arbor, MI 48109, USA.
* Correspondence: farhang@umich.edu

**Abstract:** 4D printing is a new manufacturing paradigm that combines stimuli-responsive materials, mathematics, and multi-material additive manufacturing to yield encoded multi-material structures with intelligent behavior over time. This emerging field has received growing interests from various disciplines such as space exploration, renewable energy, bioengineering, textile industry, infrastructures, soft robotics, and so on. Here, as a first attempt, we consider the energy aspect of 4D printing. By a thermodynamic analysis, we obtain the theoretical limit of energy consumption in 4D printing and prove that 4D printing can be the most energy-efficient manufacturing process. Before that, we clearly underpin 4D printing as a new manufacturing process and identify its unique attributes.

## 1. Introduction

Manufacturing industries consume about one-third (31%) of the global energy and are also responsible for approximately one-third (36%) of $CO_2$ emissions [1–4]. Energy availability and costs are the next issues in addition to environmental impacts [5,6]. Therefore, energy efficiency has been the focus of many studies recently, and its eminence has been highlighted more than ever [7,8].

On the other hand, 4D printing [9–14] provides a situation for using random (free) energy to make non-random structures [9]. This goal is achieved by utilizing and encoding smart (stimuli-responsive) materials in multi-material structures. 4D printing can revolutionize manufacturing and construction industries [9]. This emerging field has received growing interests from various disciplines such as space exploration [15–17], renewable energy [18,19], bioengineering [20], textile industry [21,22], soft robotics [23,24], electrical circuits [25], metamaterials [26–28], kirigami [29], origami [30], supercapacitors and batteries [31], infrastructures [9], and so on.

In the following, first, we underpin 4D printing as a new manufacturing process and identify its unique attributes. Second, we organize various concepts of assembly unveiled over time. Third, we introduce principles of self-assembly at manufacturing scale as it is in the early stage of development and convergence of ideas onto the right direction is required. Fourth, we obtain the theoretical limit of minimum energy consumption in manufacturing that can be approached by 4D printing.

## 2. 4D printing as a new manufacturing process with unique attributes

3D printing (additive manufacturing) is a well-known manufacturing process with its unique attributes. Now, 4D printing needs to be clearly defined and described as a new manufacturing process and its unique attributes should also be proved.

By analyzing natural shape morphing materials and structures [32], in addition to stimuli and stimuli-responsive materials, one other thing is observed that is encoded anisotropy [32]. To enable complicated shape-shifting behaviors required for accomplishing various tasks in natural structures such as the pinecone, nature programs a specific arrangement of active and passive elements [32]. This encoded anisotropy is required to direct the response into the desired direction [32]. Now, 4D printing is a good paradigm to meet this type of encoding in synthetic shape morphing structures. By considering the aforementioned point in natural shape morphing structures and based on the 4D printing concepts discussed in our previous review article [14], as well as some of the related works [9–12], we underpin 4D printing as a new manufacturing process as shown in Figure 1. We also identify attributes of 4D printing as a new manufacturing process as Figure 2. Almost all applications enabled by 4D printing can be categorized into self-assembly, self-adaptability, and self-repair [14] that we call them here as "3S 4D printing applications" (Figure 2). 4D printing conserves the advantages of 3D printing and further adds new features (Figure 2). The "complexity-free geometry" attribute was introduced as the unique feature of structures made by 3D printing [33–36]. Here, we introduce "complexity-free geometry change" as the unique attribute of 4D printed structures. One of the key goals of 3D printing is movement from form to functionality [37]. 4D printing goes further and provides multi-functionality. Furthermore, 4D printing possesses the material-saving characteristic arising from the general advantages of 3D printing and further adds the energy-saving trait (that is the focus of this study).

**Proposal for future:** it should be highlighted that future "4D printers" should possess an integrated inverse mathematics (as a software/hardware added to the current multi-material 3D printers) to predict the shape-shifting behaviors for various (or categories of) materials and stimuli. The term "4D printer" has already been mentioned in various studies in the literature of 4D printing. However, "4D printer" is *not* simply achieved by extending a single-material 3D printer to a multi-material 3D printer, or by combining several printing techniques (e.g., FDM and inkjet) in one 3D printer. We should say that *"4D printer" should be able to analyze and predict the "4$^{th}$ D"*. To achieve a 4D printer, an "intelligent head" should be developed and added to the current printers (Figure 3). This head (as an integrated software/hardware added to the existing multi-material 3D printers) should be able to analyze and predict the 4$^{th}$ D. It should be able to predict the appropriate arrangement of active and passive materials (an encoded anisotropy) for the desired evolution after printing. As we elaborated in our previous work [14], 4D printing mathematics is a link between four main factors: printing path (arrangement of active and passive voxels), desired shape after printing, stimulus properties, and materials properties. 4D printing mathematics is required to predict the shape-shifting behavior after printing over time, prevent internal collisions, and decrease or even eliminate trial-and-error tests for getting the desired shape-shifting [14]. Currently, 4D printing process utilizes the inverse mathematical modeling in an offline manner (passively), as seen in Figure 1. However, the inverse mathematical modeling can systematically be incorporated into current 3D printers to yield 4D printers that can analyze and predict the 4$^{th}$ D.

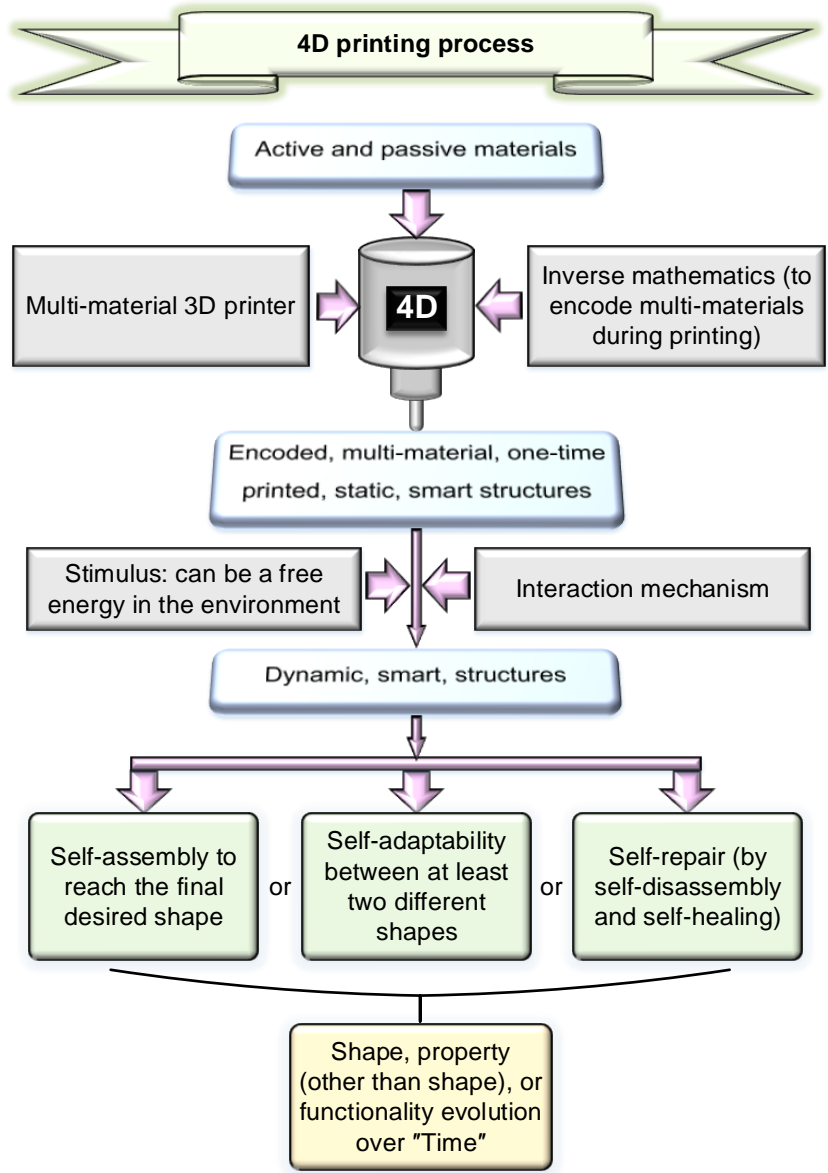

**Figure 1.** 4D printing process.

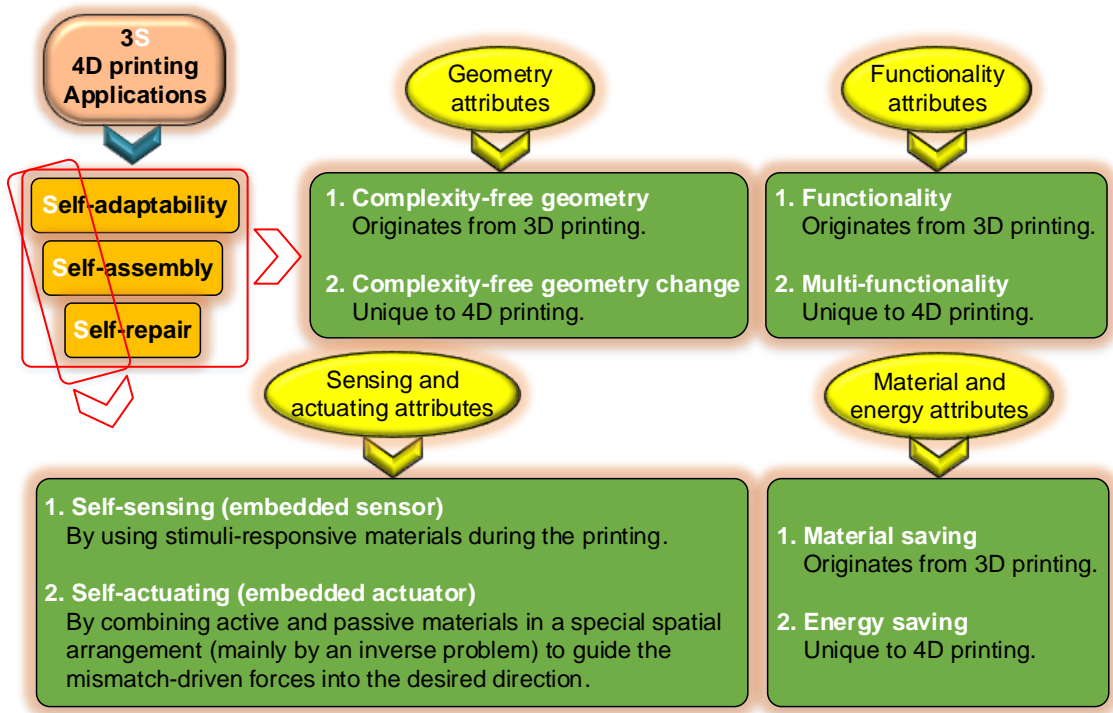

**Figure 2.** "3S 4D printing applications" and 4D printing attributes.

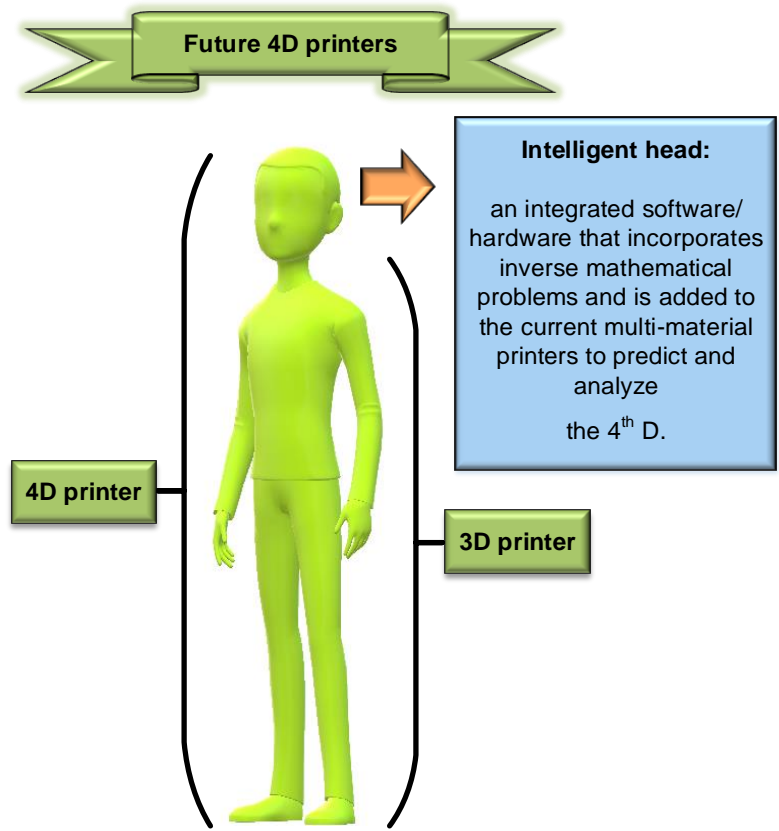

**Figure 3.** Future 4D printers. The term "4D printer" has already been mentioned in some studies. However, "4D printer" is *not* simply achieved by extending a single-material printer to a multi-material printer, or by combining several printing techniques (e.g., FDM and inkjet) in one 3D printer. "4D printer" should be able to analyze and predict the "4th D". Thus, to achieve a 4D printer, an "intelligent head" (i.e., an integrated software/hardware that incorporates inverse mathematical problems of Figure 1) should be developed and added to the current multi-material 3D printers.

## 3. (R)evolution of assembly concept in manufacturing

In this section, we analyze the literature [38–50,9–14] and organize various concepts of assembly in manufacturing as seen in Figure 4.

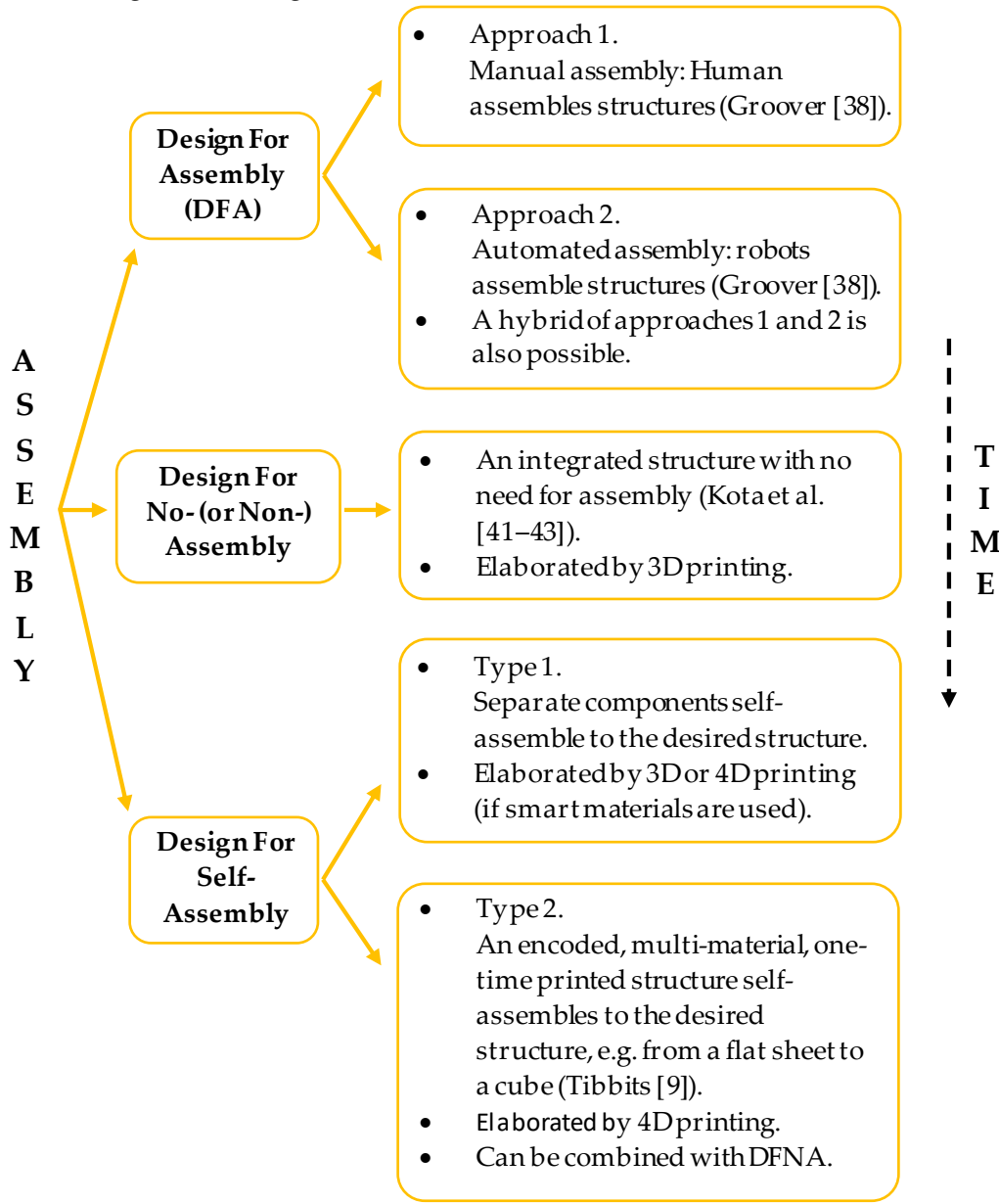

**Figure 4.** The (r)evolution of assembly concept in manufacturing.

## 4. Principles of self-assembly at manufacturing scale

Self-assembly is a process with low or no waste [51] and covers from the molecular- to the galactic-scale applications [52]. Even though the self-assembly has been initiated at the molecular scale [52], it can be more important for making products and structures at nano, meso, and macro scales [53] that we call them here as manufacturing scale. By analyzing the principles of self-assembly at the molecular scale [53], the concepts of assembly at manufacturing scale [40], and 4D printing concepts [14], we introduce principles of self-assembly at manufacturing scale (covering both types I and II discussed in Figure 4).

- *Smart components*

Physical components are needed for making structures through self-assembly [53]. In addition, the components (whether separate components in self-assembly type I, or the multi-component structure in self-assembly type II) should have smartness to self-assemble on their own. This smartness can be embedded into structures by using smart materials and enabled by stimuli.

- *Feature modeling*

When two components (in self-assembly type I) or two parts of one multi-component structure (in self-assembly type II) are planned to mate, at least, two features, each of which located on one of the two parts are needed (Figure 5(b)) [40]. To have a systematic analysis for various purposes such as achieving a specific distance or angle between two points of the assembled structure, the assembling parts, and their features should be modeled by coordinate frames. After assigning coordinate frames to each part (e.g., $A$) and its feature (e.g., $F_A$), the assembly process can be modeled by matrix transformations (e.g., $T_{A-F_A}$) to analyze translation and orientation of assembling parts (Figure 5(b)) [40]. These features can be elaborated by the additive manufacturing.

- *Variation analysis*

In practice, various errors such as mislocations of features on parts, misshapen features, or mating errors between features can occur [40]. Therefore, variation analysis, (in which $T' = T\, dT$ is the central equation [40]), should be implemented in conjunction with the feature modeling (Figure 5(c)) [40].

- *Constraint modeling*

Each part in space can have three linear ($v_x, v_y, v_z$) and three angular ($\omega_x, \omega_y, \omega_z$) velocities (motions). If a part is constrained in a linear/angular direction, it can resist a force ($f$)/moment ($m$) in that direction. Otherwise, it has a motion in that direction. Screw Theory [54–56] is a tool to implement constraint/motion analysis by performing reciprocal, union, and intersection operations on Twist and Wrench matrices of features (Figure 5(d)) [40]. This analysis is required to identify under-constraint, over-constraint, and properly-constraint parts in an assembly and devise appropriate decisions [40].

- *Sequence analysis*

Theoretically, several assembly sequences can be realized in an assembly, but not all of them are feasible. First, the feasible assembly sequences should be obtained by using one of the existing methodologies such as AND/OR graph developed by De Mello and Sanderson (Figure 5(e)) [57,40]. Then, the best assembly sequence(s) can be chosen among the feasible ones based on technical and business criteria [40]. The final thing that should be considered in assembly modeling is the Datum Flow Chain (DFC) for extra filtering of assembly sequences and a systematic analysis of Key Characteristics (KCs) (Figure 5(e)) [40].

- *Stimulus*

A stimulus such as heat, light, water, humidity, etc., which can be a free resource in the environment is needed to trigger the self-assembly process.

- *Interactions*

In a self-assembling system, interactions between separate components [53] in self-assembly type I or between active and passive components (materials) in the multi-component structure in self-assembly type II, as well as interactions between stimulus and component(s) should be analyzed.

- *Mobility*

Mobility under the right stimulus is required in self-assembling systems. At large scales, gravity and friction forces should also be taken into consideration [53].

- *Reversibility*

"Self-assembly" is different from "formation" [52]. Reversibility is one of the primary requirements for a self-assembly process that also allows for adjustability and error-correct [53].

The initial 4D printed structures [58,11,10] were generally reversible. However, their reversibility could be purposely prevented by devising some self-locking mechanisms. These self-locking mechanisms are specific arrangements of active and passive materials that are planned in multi-material structures to intentionally prevent the reversibility when the stimulus is off. Furthermore, in future, 4D printing can freely provide both the reversible and irreversible structures [59], depending on needs.

- *Inverse problem*

In the self-assembly at manufacturing scale, there is no human or robot to put the components in the right positions (in self-assembly type I) or to bring various parts of a multi-component, one-time printed structure into correct locations (in self-assembly type II). Therefore, an inverse problem should be modeled by considering all the previous principles to achieve the final desired structure, accurately.

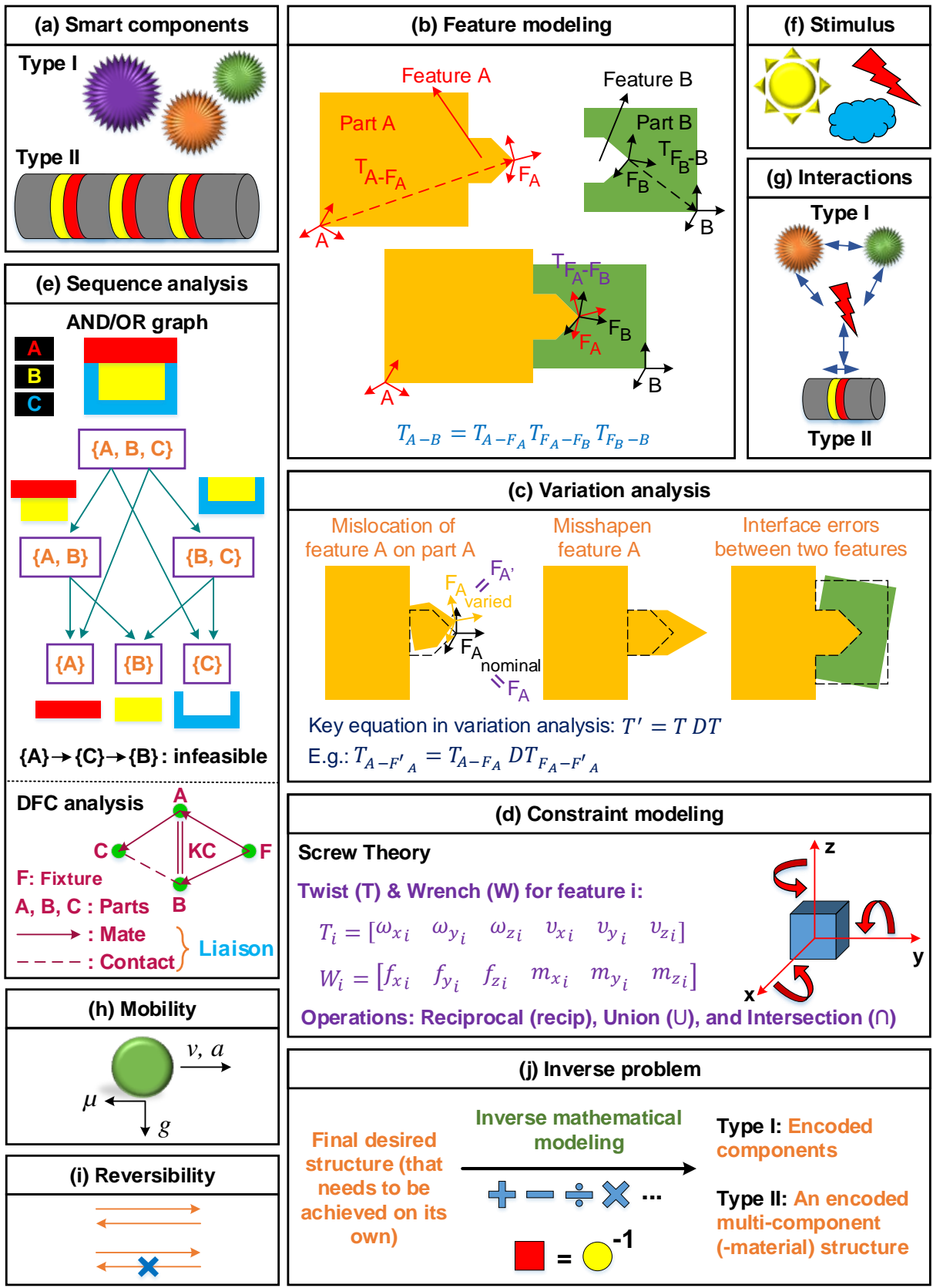

**Figure 5.** The principles of self-assembly at manufacturing scale. By analyzing the principles of self-assembly at the molecular scale [53], the concepts of assembly at manufacturing scale [40], and 4D printing concepts [14], we introduce principles of self-assembly at manufacturing scale.

## 5. Energy aspect of 4D printing as a new process for self-assembly at manufacturing scale

A manufacturing process, in the most general form, can be modeled as Figure 6 [60–64]. There are three types of energy transfer mechanisms between a system and its surroundings: heat, work, and mass flow (in fact, mass is energy, and mass flow is present in open thermodynamic systems) [65]. The energy transfer between a system and its surroundings causes entropy transfer between them so that $\dot{S}_{heat} = \frac{\dot{Q}}{T}$, $\dot{S}_{work} = 0$, and $\dot{S}_{mass} = \dot{m}s$ (there is no entropy transfer by work) [65]. Energy is conserved (i.e., it cannot be destroyed or generated), while entropy can be generated [65]. The first and second laws of thermodynamics deal with energy and entropy, respectively, and in the most general forms are [65]:

$$\begin{cases} \text{Energy balance: } E_{in} - E_{out} = \Delta E_{sys} \xrightarrow{\text{rate form}} \dot{E}_{in} - \dot{E}_{out} = \frac{dE_{sys}}{dt} \\ \text{Entropy balance: } S_{in} - S_{out} + S_{gen} = \Delta S_{sys} \xrightarrow{\text{rate form}} \dot{S}_{in} - \dot{S}_{out} + \dot{S}_{gen} = \frac{dS_{sys}}{dt} \end{cases} \quad (1)$$

By ignoring kinetic and potential energies (and other macroscopic forms of energy), the two thermodynamics laws are [66]

$$\begin{cases} \text{Energy balance: } \dot{Q}_{in} - \dot{Q}_{out} + \dot{W}_{in} - \dot{W}_{out} + \sum_{in} \dot{m}h - \sum_{out} \dot{m}h = \frac{dE_{sys}}{dt} \\ \text{Entropy balance: } \sum_H \frac{\dot{Q}_{in}}{T_H} - \sum_L \frac{\dot{Q}_{out}}{T_L} + \sum_{in} \dot{m}s - \sum_{out} \dot{m}s + \dot{S}_{gen} = \frac{dS_{sys}}{dt} \end{cases} \quad (2)$$

Here, we adopted subscripts $H$ and $L$ to indicate high and low temperatures, respectively (Figure 6).

By applying equation (2) to a general manufacturing process (Figure 6) operating under steady conditions ($\frac{d}{dt} = 0$), we have

$$\begin{cases} \text{Energy balance: } \dot{Q}_{in} - \dot{Q}_{out} + \dot{W}_{in} - \dot{W}_{out} + (\dot{m}h)_{in} - (\dot{m}h)_{out} = 0 \\ \text{Entropy balance: } \frac{\dot{Q}_{in}}{T_H} - \frac{\dot{Q}_{out}}{T_L} + (\dot{m}s)_{in} - (\dot{m}s)_{out} + \dot{S}_{gen} = 0 \end{cases} \quad (3)$$

By multiplying both sides of the entropy balance equation by $T_L$ and equating the left-hand sides of the energy balance and the resulting entropy balance equation, we arrive at

$$\dot{W}_{in} = \dot{W}_{out} + \left(\frac{T_L}{T_H} - 1\right)\dot{Q}_{in} + (\dot{m}h)_{out} - (\dot{m}h)_{in} - T_L[(\dot{m}s)_{out} - (\dot{m}s)_{in}] + T_L\dot{S}_{gen}. \quad (4)$$

Equation (4) is a more general form of the required input work than that obtained in the literature [63,64], which further considers the output work. Up to here, we followed the same approach devised by Gutowski et al. [61–64] and further extended their required-input-work equation.

Now, the required input power (energy rate) would be

$$\dot{E}_{in} = \dot{W}_{in} + \dot{Q}_{in} \Rightarrow$$

$$\dot{E}_{in} = \dot{W}_{out} + \frac{T_L}{T_H}\dot{Q}_{in} + (\dot{m}h)_{out} - (\dot{m}h)_{in} - T_L[(\dot{m}s)_{out} - (\dot{m}s)_{in}] + T_L\dot{S}_{gen}. \quad (5)$$

This equation can also be written in non-rate form as

$$E_{in} = W_{out} + \frac{T_L}{T_H}Q_{in} + (mh)_{out} - (mh)_{in} - T_L[(ms)_{out} - (ms)_{in}] + T_L S_{gen}. \quad (6)$$

Equation (6) is a general equation that gives the required input energy for any manufacturing process.

Now, for 4D printing as a manufacturing process that enables self-assembly at the manufacturing scale, this equation can be further analyzed and simplified. Let us consider the following key-points:

- **Key-point 1.** Self-assembly is a spontaneous and reversible process [67–69,51–53]. For such a process, the central thermodynamic equation is $\Delta G_{process} = 0$ [67,65].
- **Key-point 2.** In a reversible process, the system is in thermodynamic equilibrium with its surroundings. One of the necessities of thermodynamic equilibrium is thermal equilibrium. When two bodies are in thermal equilibrium, their temperatures are the same. Therefore, during a reversible process, $T_{sys} \approx T_{surr}$ [70,71,65].
- **Key-point 3.** For a reversible process, $S_{gen} = 0$ [65].

By considering Key-point 2, for any manufacturing process in the reversible condition, equation (6) can be written as (we emphasize that from equation (1), the convention is $\Delta \odot = \odot_{in} - \odot_{out}$):

$$E_{in} = W_{out} + \frac{T_L}{T_H} Q_{in} - \Delta G + T_L S_{gen}, \qquad (7)$$

where $\Delta G = \Delta H - T\Delta S$ and $G$ is the Gibbs free energy. Then, by using key-points 1 and 3, we have the following equation for 4D printing that enables self-assembly at manufacturing scale:

$$E_{in} = W_{out} + \frac{T_L}{T_H} Q_{in}. \qquad (8)$$

Equation (8) is the minimum theoretical limit of required input energy for 4D printing as a new process that enables self-assembly at manufacturing scale. In addition, generally, $\Delta G \leq 0$ and $S_{gen} \geq 0$ and thus both terms $-\Delta G$ and $T_L S_{gen}$ are positive or zero ($\geq 0$). Therefore, by comparing equations (7) and (8), it can be concluded that 4D printing can have the minimum energy consumption among various manufacturing processes.

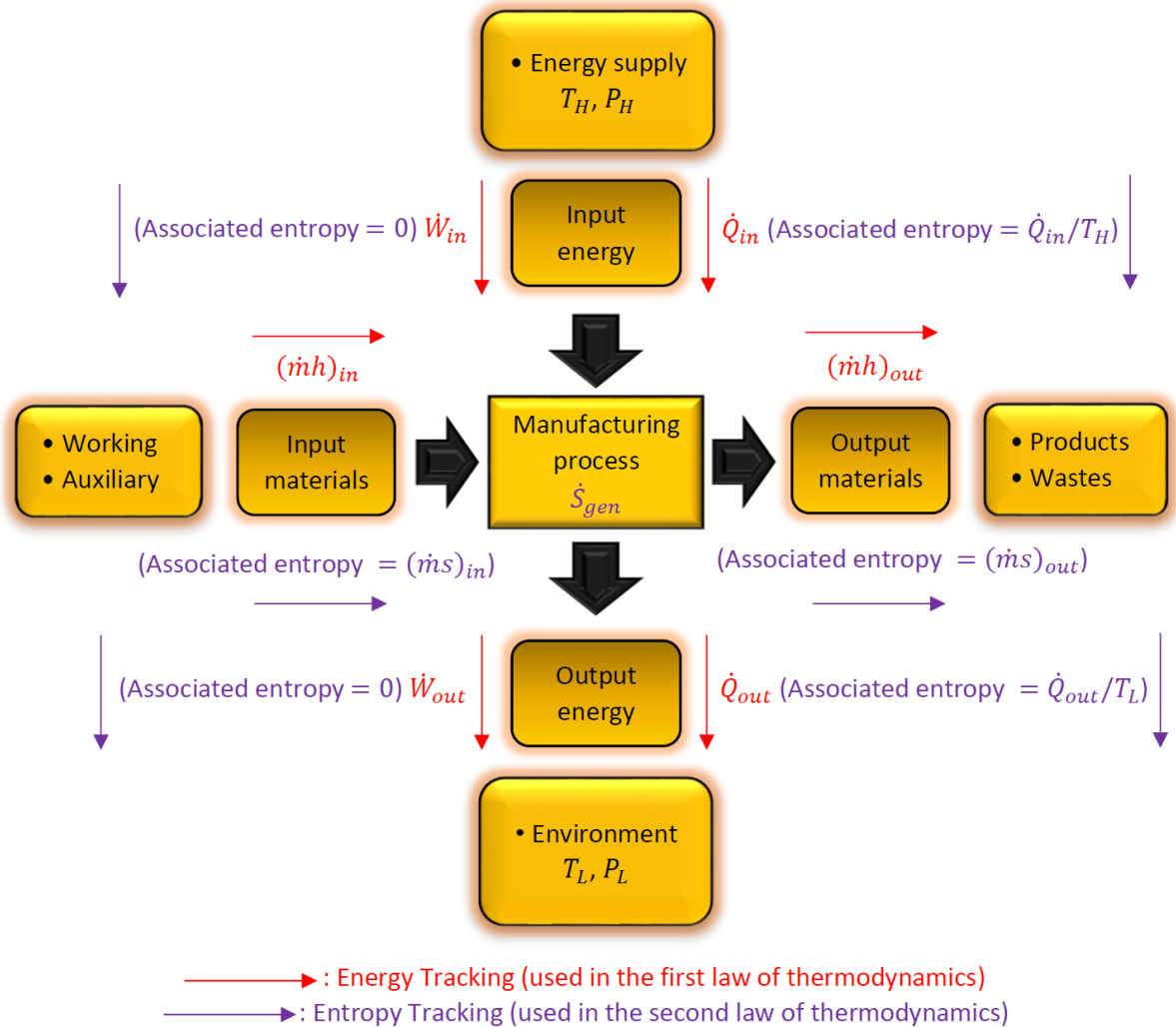

**Figure 6.** A manufacturing process in the most general thermodynamic model (this figure has been drawn based on the concepts in [60–64]). The energy and entropy flows have also been illustrated.

## 6. Conclusion

We derived the theoretical limit of minimum energy consumption in 4D printing as a manufacturing process and proved that 4D printing could be the most energy-efficient process among various manufacturing processes. This minimum energy consumption limit in manufacturing obtained here can be *approached* by 4D printing process and future 4D printers. It does not necessarily mean that this limit is practically achieved in 4D printing processes. One of the main reasons is that, currently, in 4D printing, although the self-assembly process can be triggered by environmental free energy, fabrication of the initial self-assembling components requires electrical energy for running the printers. Nevertheless, future 4D printers may somehow incorporate environmental free energy for the whole manufacturing process (that is, fabrication of the self-assembling components and then self-assembly of them). Sadi Carnot worked on the energy efficiency of "heat engines" and the Carnot cycle gives the theoretical limit in heat engines. Here, we worked on the energy efficiency of "manufacturing processes" and obtained the theoretical limit of minimum energy consumption in manufacturing that can be approached by 4D printing. In this study, we have also clearly underpinned 4D printing as a new manufacturing process with unique attributes.

**Acknowledgments:** F.M. wants to thank the graduate-level course "Assembly Modeling for Design and Manufacturing" taught by Prof. Kazuhiro Saitou at the University of Michigan-Ann Arbor, for which F.M. served as GSI (Graduate Student Instructor). This course was helpful to analyze the concepts of "assembly" at manufacturing scale.